# First-Principles Study of Hydrogen Behaviors in α-Pu$_2$O$_3$

Le Zhang[1], Bo Sun[1,*], Qili Zhang[1], Haifeng Liu[1], Kezhao Liu[2] and Haifeng Song[1,3,*]

[1]*Institute of Applied Physics and Computational Mathematics, No.2 of Fenghao East Road, Beijing 100094, People's Republic of China*

[2]*Science and Technology on Surface Physics and Chemistry Laboratory, Mianyang 621908, People's Republic of China*

[3]*CAEP Software Center for High Performance Numerical Simulation, No. 6 of Huayuan Road, Beijing 100088, People's Republic of China*

E-mail: sun_bo@iapcm.ac.cn; song_haifeng@iapcm.ac.cn

**Abstract**

The in-depth understanding of hydrogen permeation through plutonium-oxide overlayers is the prerequisite to evaluate the complex hydriding induction period of Pu. In this work, the incorporation, diffusion and dissolution of hydrogen in α-$Pu_2O_3$ are investigated by the first-principles calculations and *ab initio* thermodynamic method based on DFT+$U$ and DFT-D3 schemes. Our study reveals that the hydrogen incorporation is endothermic and the separated H atoms prefer to recombine as $H_2$ molecules rather than reacting with α-$Pu_2O_3$. The H and $H_2$ diffusion are both feasible, generally, H will recombine first as $H_2$ and then migrate. Both pressure $P_{H2}$ and temperature can promote the hydrogen dissolution in α-$Pu_2O_3$. The single $H_2$ molecule incorporation and (H+$H_2$) mixed dissolution will successively appear when increasing $P_{H2}$. Compared to $PuO_2$, this work indicates that Pu sesquioxide is hardly reduced by hydrogen, but the porous α-$Pu_2O_3$ facilitates hydrogen transport in Pu oxide layers. We presents the microscopic picture of hydrogen behaviors in the defect-free α-$Pu_2O_3$, which could shed some light on the study of the hydriding induction period of Pu.

## 1. Introduction

Plutonium (Pu) is strategically important in nuclear explosives and nuclear energy supply. The chemical and physical properties of Pu are very sensitive to environment due to its complex 5f states [1-4]. As a result, the surface oxidation of Pu is unavoidable during its long-term storage, handling or preparation [5]. In air, Pu surface is quickly covered by oxide layers, mainly consisting of the $PuO_2$ outer-layer and the $\alpha\text{-}Pu_2O_3$ inter-layer [6]. During long-term storage or in ultra-high vacuum environment, $PuO_2$ will reduce to $\alpha\text{-}Pu_2O_3$ spontaneously [7]. The hydriding corrosion of Pu is catastrophic, because Pu-hydride can violently accelerate oxidation to pyrophoricity of Pu [5-6]. According to the diffusion barrier model, Pu-hydriding process can be divided into four periods: induction, nucleation or acceleration, bulk hydriding and termination [8-10]. The sequence of main steps in the induction period is usually proposed as: (i) $H_2$ physisorption and dissociative chemisorption on Pu-oxide surface; (ii) H penetration into Pu-oxide subsurface; (iii) H diffusion and dissolution in $PuO_2/\alpha\text{-}Pu_2O_3$ overlayer; (iv) H transfer across $\alpha\text{-}Pu_2O_3/Pu$ interface; (v) H dissolution and diffusion in Pu matrix; and (vi) H accumulation to saturation and preliminary Pu-hydride nucleation [5,11].

Since hydrogen reacts violently with Pu metal, above (i), (ii) and (iii) steps involved with the interactions between hydrogen and Pu-oxide overlayers could be rate-limiting steps of Pu-hydriding process [5]. Actually, many experiments have been devoted to investigating the influence of Pu-oxide overlayers on Pu-hydriding. Early experiments have revealed that $PuO_2$ works as the protective screen against Pu-hydriding and results in the induction period, but $\alpha\text{-}Pu_2O_3$ can not effectively resist hydrogen attack [5,12-13]. Haschke *et al.* once inferred that $\alpha\text{-}Pu_2O_3$ can catalyze $H_2$ dissociation and act as a medium for H transport to Pu matrix [5]. Dinh *et al.* further compared the hydriding rate of Pu, $PuO_2$, $\alpha\text{-}Pu_2O_3$ samples to investigating the role of $\alpha\text{-}Pu_2O_3$ in Pu-hydriding, and validated that the induction period of Pu covered with $\alpha\text{-}Pu_2O_3$ is similar to Pu and much shorter than Pu covered by $PuO_2$ [13]. Meanwhile, they proposed that H solubility in $PuO_2$ is much lower than the solubility limit in Pu, thus $PuO_2$ has the blocking effect on Pu-hydriding, and α-$Pu_2O_3$ mainly promotes the nucleation process [13-14]. They also pointed out that more experimental efforts are expected to clarify such unclear micro-mechanisms as the solubility and diffusion of hydrogen species in Pu-oxide overlayers.

Due to the complex Pu-oxide overlayer, the radioactivity and toxicity of Pu, experimental researches on micro-mechanisms of above-mentioned rate-limiting steps meet with so many challenges that the corresponding atomic-level theoretical studies have been calling for. In fact, several theoretical studies have been performed to investigate the hydrogen behaviors in the (i) and (ii) steps of induction period. Our previous work [15] revealed that defect-free $PuO_2$ (111) surface can suppress (i) and (ii) steps. When defect-free α-$Pu_2O_3$ (111) exposed, the micro-pictures of (i) and (ii) steps are quite different, namely, $H_2$ penetrates into α-$Pu_2O_3$ with a much lower barrier than collision-dissociation barrier of $H_2$ on both $PuO_2$ and α-$Pu_2O_3$(111) surfaces [15]. H. L. Yu *et al*. showed that even on the $PuO_2$ (110) surface with Pu cations exposed, there is also a certain energy barrier for $H_2$ dissociation although the chemisorption of H is much more exothermic [16], and the barrier of step (ii) rises to 2.15 eV [17]. These theoretical studies have obtained the detailed micro-pictures on surface hydriding process consisting of (i) and (ii) steps. However, there are considerably fewer studies on the (iii) step of hydrogen behaviors in bulk Pu-oxides. In order to achieve in-depth understanding of Pu-hydriding induction period and establish a predictive model, the micro-mechanism of hydrogen transport in $PuO_2$ and α-$Pu_2O_3$ is critical. However, because α-$Pu_2O_3$ is the last screen for Pu layer and must be present at oxide-metal interface [5], current experiments can not directly detect the microscopic behaviors of hydrogen in α-$Pu_2O_3$ layer.

The first-principles calculations and *ab initio* atomistic thermodynamic method are performed to studying the hydrogen behaviors in the defect-free α-$Pu_2O_3$ layer, aiming at clarifying the micro picture of hydrogen interaction with α-$Pu_2O_3$ matrix: (a) the stable incorporation states of hydrogen in α-$Pu_2O_3$; (b) the diffusion path and mechanism of hydrogen in α-$Pu_2O_3$; and (c) the temperature and pressure dependent dissolution mechanism of hydrogen in α-Pu$_2$O$_3$.

## 2. Methods

The first-principles calculations were performed by Vienna *ab initio* simulation package (VASP) [18]. The exchange-correlation interaction of electrons was described by Perdew-Burke-Ernzerhof (PBE) of generalized gradient approximation (GGA) [19-20]. The 6s$^2$7s$^2$6p$^6$6d$^2$5f$^4$ and 2s$^2$2p$^4$ electrons of Pu and O were considered as the valence electrons [15,21-23]. According to experimental measurements, the ground state of α-Pu$_2$O$_3$ was set to be antiferromagnetic (AFM) [24-25]. The van der Waals correlation was described by DFT-D3 method [26-28]. The Brillouin zone was sampled with a 4 × 4 × 4 Monkhorst-Pack *k*-point grid [29]. All structures were relaxed until the residual force on each atom was less than 0.01 eV Å$^{-1}$. The plane-wave cutoff energy was set to 600 eV. The strong on-site Coulomb repulsion among the Pu-5f electrons was described with the DFT+*U* scheme in the widely applied formalism of S. L. Dudarev *et al* [30]. The difference between Coulomb energy *U* and exchange energy *J* parameters, $U_{eff}$=*U-J*, was significant for the total energy functional, during calculations $U_{eff}$ was set to 4 eV.

The calculated lattice constant of α-Pu$_2$O$_3$ is 11.11 Å, which is close to experimental values (11.03 Å ~ 11.07 Å) [31-32]. The α-Pu$_2$O$_3$ crystal cell constructs with 25% oxygen vacancies along 16c (0.25, 0.25, 0.25) sites in (2×2×2) supercell of stoichiometric PuO$_2$ [15], thus the coordination environments of O and Pu are different from those in stoichiometric PuO$_2$. As shown in Fig. 1(b), Pu tetrahedron is structurally distorted, the Pu-O bond lengths range in 2.35 Å ~ 2.44 Å and Pu-O-Pu angles range in 100.04º ~ 124.20º. Pu cations in α-Pu$_2$O$_3$ are not equivalent and can be divided into two types by their bonding property, labeled as "Pu$^{Type-1}$" (black ball) and "Pu$^{Type-2}$" (blue ball) in Fig. 1(a)-(c). There are eight Pu$^{Type-1}$ cations in bulk α-Pu$_2$O$_3$, which are fixed at the face center, body center, vertex and middle of edges, respectively. The α-Pu$_2$O$_3$ crystal cell can be divided into eight equivalent cubic cages, which correspond to eight crystal cells of PuO$_2$, but a hole structure appears between adjacent cages as with oxygen vacancies. In each cage, $Pu^{Type-2}$ cations and O anions deviate from their corresponding lattice sites in $PuO_2$ crystal cell, and then constitute a ring structure (Fig. 1(d)). As shown in Fig. 1(c), the $Pu^{Type-1}$ cation connects six O anions from six rings with equal $Pu^{Type-1}$-O bond length (2.39 Å) and the $Pu^{Type-2}$ cation connects two nearest rings with $Pu^{Type-2}$-O bond length ranging from 2.35 Å to 2.44 Å.

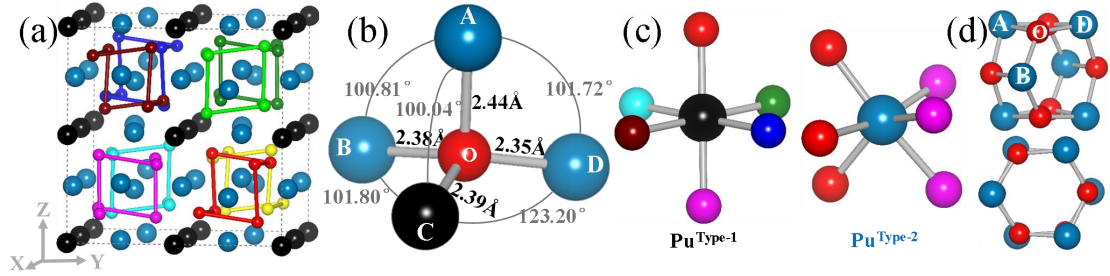

Fig. 1 (a) Calculated structure of defect-free α-$Pu_2O_3$ crystal cell, the brown, blue, cyan, magenta, red, yellow, green, dark green balls represent O anions in eight cages; (b) the distorted tetrahedron structure of four Pu cations, labeled in A, B, C, D; (c) the coordination environment of $Pu^{Type-1}$ and $Pu^{Type-2}$ cations; (d) the ring formed by $Pu^{Type-2}$ and O in each cage.

The incorporation energy $E_i$ of H in α-$Pu_2O_3$ is calculated by

$$E_i = E_{Pu2O3+xH} - E_{Pu2O3} - x/2 E_{H2} , \qquad (1)$$

Where $E_{Pu2O3+xH}$, $E_{Pu2O3}$ and $E_{H2}$ represent the total energy of H in α-$Pu_2O_3$, total energy of bulk α-$Pu_2O_3$ and $H_2$, respectively. The energy barriers are calculated by the climbing image nudged elastic band method (CI-NEB) with three images between initial and final structures [33-34]. The zero point energy is considered in the calculation of $E_i$ and energy barriers.

The "*ab initio* atomistic thermodynamic method" [35-36] is used to calculate the changes of Gibbs free energy ΔG(T,P), shown as

$$\Delta G(T,P) = (E_i - N \times \mu_{H2}(T,P))/V , \qquad (2)$$

$$\mu_{H2}(T,P) = \mu_{H2}(T,P_0)_s + k_B T \ln(P_{H2}/P_0) , \qquad (3)$$

$$\mu_{H2}(T,P_0) = \Delta H_{H2}(T,P_0) - T\Delta S_{H2}(T,P_0) , \qquad (4)$$

where $\mu_{H2}(T,P)$ is the chemical potential of $H_2$. In this work, we discuss the dissolution heat of hydrogen in bulk α-$Pu_2O_3$, V and N in equation (2) represent the volume of bulk α-$Pu_2O_3$ and number of $H_2$ molecules. Under the approximate of ideal gas, the chemical potential of $H_2$ can be written as equation (4), $k_B$ and $P_{H2}$ are the Boltzmann constant and partial pressure of $H_2$. $\Delta H_{H2}$ and $\Delta S_{H2}$ represent the changes of enthalpy and entropy of $H_2$ at standard state pressure $P_0$ with the temperature ranging from T to 0K.

## 3. Results and Discussions

### 3.1 Incorporation Properties of H and $H_2$ in α-$Pu_2O_3$

Owing to the oxygen vacancies, H atom can incorporate at more potential sites in α-$Pu_2O_3$ than in $PuO_2$. Since H tends to bind to O-anion in the distorted tetrahedron of Pu, all possible incorporation sites can be found in Fig. 1(b): site in BCD area and along AO direction (Fig. 2(a)); site in hole between two cages (Fig. 2(b)); and alike sites in ACD, ABC, and ABD areas (Fig. 2(c)). The corresponding incorporation energies are 1.53 eV, 1.72 eV and 2.17 eV, which reveal that H incorporation in α-$Pu_2O_3$ is endothermic and H prefers to bind to O-anion through BCD surface. Besides, H incorporation through BCD surface with a 0.99 Å H-O distance which slightly causes structure distortion of α-$Pu_2O_3$ with the $Pu^A$-O bond lengthening 0.52 Å.

In order to have more insight into the stable existence state of hydrogen in α-$Pu_2O_3$, we further calculate incorporation energies of two atomic H (2H) and one $H_2$ molecule. As shown in Fig .2(d), 2H absorbs at the preferred sites depicted by Fig. 2(a) with incorporation energy of 1.58 eV/H. As shown in Fig. 2(e), (f) and (g), $H_2$ locates in a $Pu^{Type-2}$-O ring, the O-vacancy site between $Pu^{Type-1}$ and $Pu^{Type-2}$-O ring, and the hole between adjacent cages, respectively. The corresponding incorporation energies of $H_2$ are 0.58 eV/$H_2$, 0.34 eV/$H_2$ and 0.89 eV/$H_2$, indicating that $H_2$ incorporation at the oxygen vacancy site is the most stable. Obviously, the endothermic incorporation of $H_2$ in α-$Pu_2O_3$ is much more stable than that of atomic H, which further verifies the difficult dissociation of $H_2$ in α-$Pu_2O_3$. As in α-$Pu_2O_3$, $Pu^{3+}$ cation can not be further reduced, but there are many holes to accommodate H$_2$ due to the porous structure.

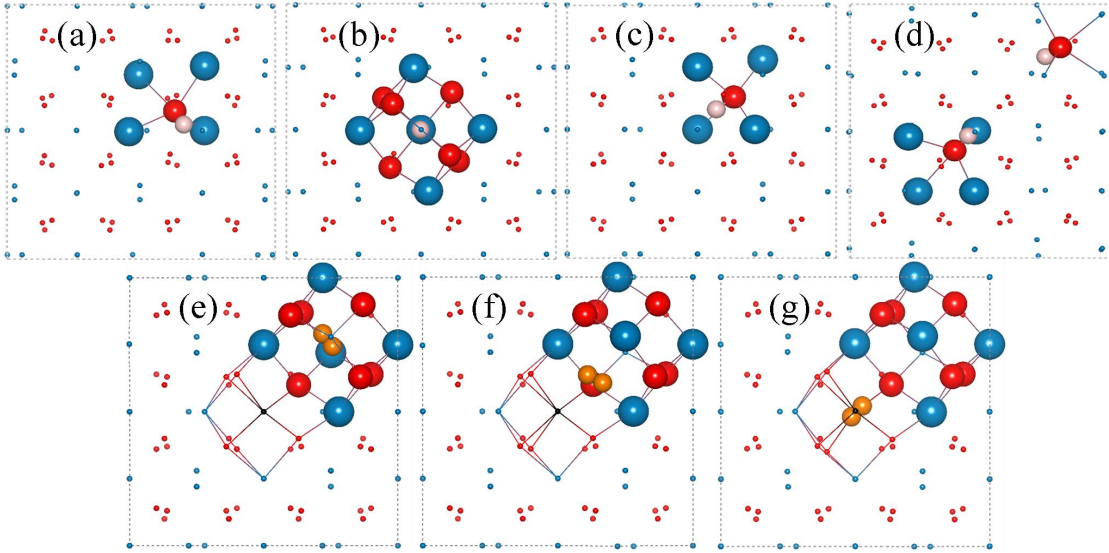

Fig. 2 (a), (b), (c) Calculated structures of one H, (d) two separate H atoms. (e), (f), (g) H$_2$ molecule incorporation in α-Pu$_2$O$_3$ depicted in the same perspective. The red and blue point, pink and orange balls represent O, Pu, H bonding with O, H$_2$, respectively.

Table 1. Incorporation energy E$_i$ of H (in eV/H) and H$_2$ (in eV/H$_2$) in Fig. 2.

|  | (a) | (b) | (c) | (d) | (e) | (f) | (g) |
|---|---|---|---|---|---|---|---|
| $U_{eff}$ =4 eV | 1.53 | 1.72 | 2.17 | 1.58 | 0.58 | 0.34 | 0.89 |
| $U_{eff}$ =3 eV | 1.50 | 1.63 | 2.08 | 1.54 | 0.60 | 0.34 | 0.91 |

DFT+$U$ studies of α-Pu$_2$O$_3$ have not formed a uniform $U_{eff}$ of 3 eV or 4 eV [37-41], in view of this, we further test the effect of $U_{eff}$ by calculating the E$_i$ of H and H$_2$ in α-Pu$_2$O$_3$ (Fig. 2) with $U_{eff}$ of 3 eV and 4 eV, respectively. As shown in Table. 1, the calculations with $U_{eff}$ of 3 eV and 4 eV come to the same conclusion that the hydrogen incorporation is endothermic and H$_2$ incorporation is much more preferred in α-Pu$_2$O$_3$. From a quantitative point of view, for the most stable incorporation state of H, the difference of the calculated E$_i$ with $U_{eff}$ = 3 eV and 4 eV is only 0.03 eV, while for H$_2$, there is no difference at all. Thus, the hydrogen behaviors in α-Pu$_2$O$_3$ can be reasonably described by $U_{eff}$ either equaling to 3 eV or 4 eV. In order to compare with our previous work [15], we perform the calculation with $U_{eff}$ of 4 eV.

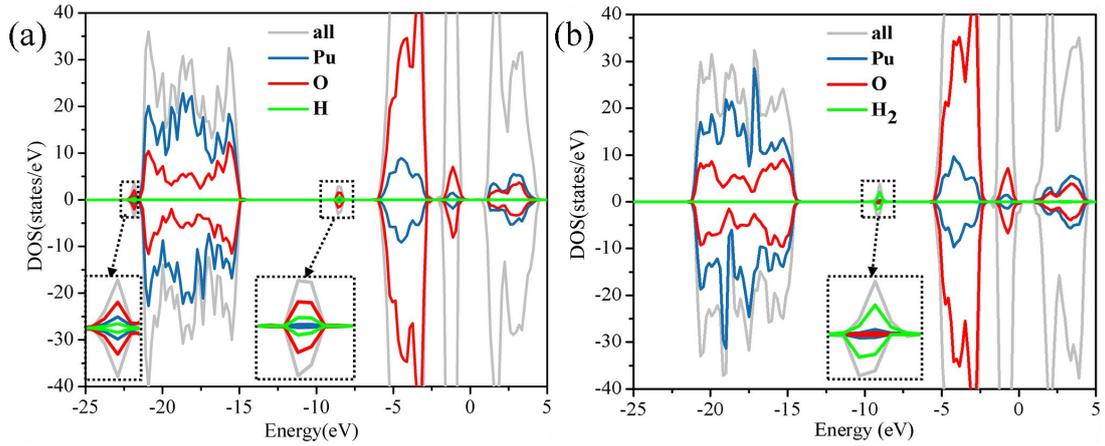

Fig. 3 (a), (b), Calculated density of states (DOS) of one H and $H_2$ incorporation in α-$Pu_2O_3$ depicted in Fig. 2(a) and Fig.2 (f).

Our previous study revealed that $H_2$ dissociation in bulk α-$Pu_2O_3$ is quite difficult [15], which can be actually described as one gas molecule of *solid-solution state* suspended in α-$Pu_2O_3$. The DOS results of the most stable incorporation state of H and $H_2$ are calculated to investigate the interactions between H, $H_2$ and α-$Pu_2O_3$. As shown in Fig. 3(a), the DOS reveals that the interaction between H and O-anion is to some extent obvious, as shown in the dot line rectangles. Whereas $H_2$ incorporation only induces one single state (the dot line rectangle), indicating $H_2$ barely interacts with α-$Pu_2O_3$. The bond-length changes of $H_2$ and the nearby Pu-O bond are negligible, which also confirm that the absorbed $H_2$ actually is like an isolated molecule.

**3.2 Diffusion Behavior of Hydrogen in α-$Pu_2O_3$**

Taking the stable H incorporation site (Fig. 2(a)) as the starting point, we mainly focus on the diffusion behavior of H, and clarify the diffusion mechanism in α-$Pu_2O_3$. Here, we propose three diffusion routes of H in α-$Pu_2O_3$, corresponding to three kinds of diffusion mechanisms. Fig. 4(a) and (b) show two possible diffusion mechanisms for H (Route 1 and Route 2), and Fig. 4(c) shows the diffusion mechanism of $H_2$ in α-$Pu_2O_3$ (Route 3).

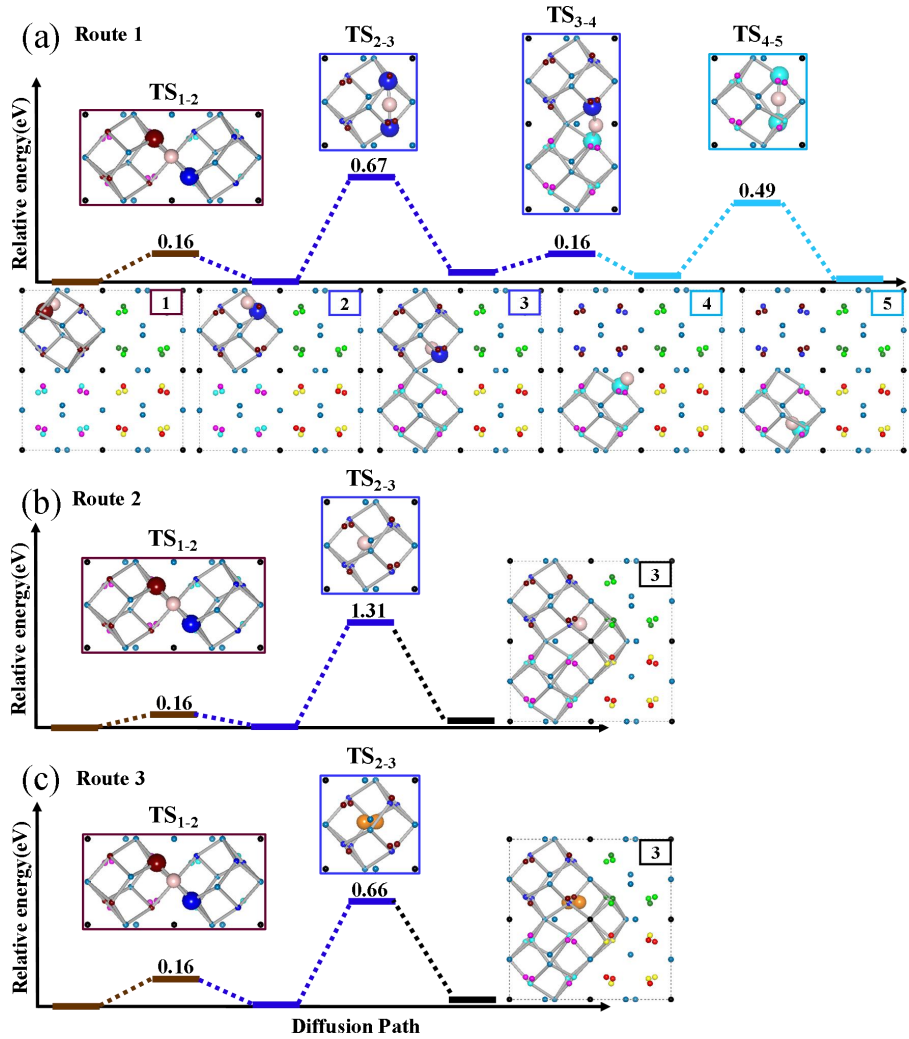

Fig. 4 Calculated H and H$_2$ diffusion path and energy barriers of (a) Route 1, (b) Route 2, and (c) Route 3 in α-Pu$_2$O$_3$. The lines in different colors indicate H diffusion in different O rings shown in Fig. 1(a). The initial and final H incorporation sites are labeled in 'n' and 'n+1', and transition states between the n-th and (n+1)-th sites are labeled as 'TS$_{n-n+1}$'.

Fig. 4(a) shows the diffusion of one H clinging to O-anions in every step. According to the calculated diffusion barriers, H diffusion in the same cage (site 2→3 and 4→5) is much more difficult than H migration between two neighboring cages (site 1→2 and 3→4). The barrier for the rate-limiting step of 2→3 is 0.67 eV and the barrier of 4→5 is 0.49 eV, whereas H diffuses between different cages with the same barrier of 0.16 eV.

Fig. 4(b) indicates that H migrates from one cage to another cage via the oxygen vacancies, which maybe a short cut, but H needs to jump into the hole depicted by site 3. Unexpectedly, in Route 2 the energy barrier for H jumping into the hole is calculated to be 1.31 eV, which is about 0.64 eV higher than the rate-limiting step in Route 1. Since the diffusion kinetics of H in α-$Pu_2O_3$ is mainly determined by the barrier, H diffusion along Route 2 is estimated as an extremely rare event.

As shown in Fig. 4(c), the third diffusion mechanism of H in α-$Pu_2O_3$ is expected to be that atomic H first recombines to $H_2$ and then diffusion along oxygen vacancies. Here, we initially put two H atoms on sites 1 and site 2 (as shown in Fig. 4(a)) with a distance of 1.81 Å, after relaxation they recombine to $H_2$ spontaneously. Then the newly formed $H_2$ will first diffuse through oxygen vacancies with an barrier of 0.66 eV and step into the $H_2$ diffusion path depicted by our previous study [15].

Between adjacent cages (oxygen vacancy site depicted in Fig.1), the energy barrier of H diffusion from site 1 to site 2 is only 0.16 eV which is 0.51 eV lower than H diffusion from site 2 to site 3, and H can recombine to $H_2$ spontaneously (in Route 3). Thus, abundant hydrogen will accumulate and recombine to $H_2$ at oxygen vacancy sites. Besides, Pu-oxide overlayers are very complicated because of the ubiquitous defects [13] and micro-cracks [14], which could act as the conduits for hydrogen transport and further shorten the hydriding induction time of Pu covered by α-$Pu_2O_3$. In the subsequent works, we will investigate the possible effects of various defects on hydrogen behaviors in α-$Pu_2O_3$.

### 3.3 Dissolution Mechanism of Hydrogen in α-$Pu_2O_3$

In this subsection, we mainly discuss the dissolution mechanism of H atoms and $H_2$ molecules in α-$Pu_2O_3$. According to the incorporation and diffusion properties of hydrogen, we propose three kinds of dissolution mechanisms named by **M1**, **M2** and **M3**.

**M1:** Because the formation of $H_2$ in hole structure is barrierless, here we just consider that several $H_2$ molecules successively penetrate into one cage of α-$Pu_2O_3$ (Fig. 5(a)-(c)). Our calculation shows that one $H_2$ molecule will escape out when the fourth H$_2$ penetrates into the same cage.

**M2:** H$_2$ molecules dissolve uniformly in eight cages, and the upper limit amount of H$_2$ dissolution in one cage is three. So that there are 8, 16 and 24 H$_2$ molecules dissolved in Fig. 5(d), (e) and (f), respectively.

**M3:** Beginning with the polyatomic incorporation states in each cage of α-Pu$_2$O$_3$, we aim at searching out the possible mixture dissolution states of H and H$_2$. Here, we consider 16, 32 and 48 H atoms dissolved uniformly in eight cages of α-Pu$_2$O$_3$. In Fig. 5(g)-(i), after relaxation, H atoms are in three types of states: H$_2$ molecules in hole (orange balls), H atoms in hole (purple balls), and H atoms bonding with O anions (pink balls).

As shown in Fig. 5, the lattice structure of α-Pu$_2$O$_3$ will be gradually distorted with the increasing of hydrogen. And the dissolved hydrogen prefers distribute in oxygen vacancies sites first. As shown in Table 2, three dissolution mechanisms are all endothermic, among which M1 is the most preferable mechanism because of the lowest incorporation energy, and then M3 is the second possible dissolution mechanism.

Table 2. Calculated incorporation energy (E$_i$, eV/H) of H dissolution in α-Pu$_2$O$_3$ for Fig. 4 and u$_{H2}$ at ΔG=0 for Fig. 5.

|  | (a) | (b) | (c) | (d) | (e) | (f) | (g) | (h) | (i) |
| --- | --- | --- | --- | --- | --- | --- | --- | --- | --- |
| E$_i$ | 0.34 | 1.06 | 1.35 | 16.96 | 24.00 | 32.40 | 8.80 | 11.52 | 21.12 |
| u$_{H2|ΔG=0}$ | 0.34 | 0.53 | 0.45 | 2.11 | 1.50 | 1.35 | 1.10 | 0.72 | 0.88 |

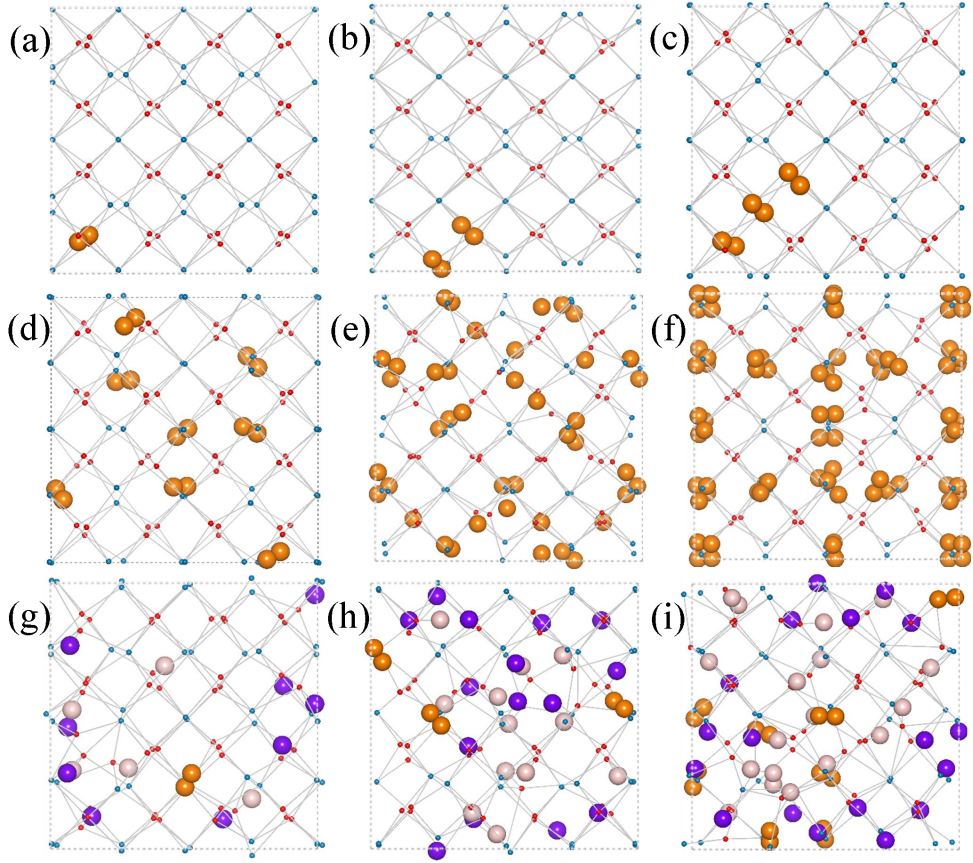

Fig. 5 Calculated structures of three penetration mechanisms: (a)-(c) M1; (d)-(f) M2 and (g)-(i) M3. The blue and red points represent Pu and O ions, the orange, pink and purple balls represent $H_2$, H bonding with O anions and H in hole, respectively.

Since the incorporation and dissolution of hydrogen in α-$Pu_2O_3$ are all endothermic, we further employ the *ab initio* thermodynamic method [35-36] to reveal the influence of external pressure of $H_2$ ($P_{H2}$) and temperature on hydrogen dissolution mechanism. The changes of Gibbs free energy (ΔG) are calculated by equation (2), the entropy and enthalpy of $H_2$ molecule in equation (4) are searched in thermochemical tables. For reference $P_0$=1 bar, the $\mu_{H2}(T,P_0)$ equals to -0.32 eV (300K) and -0.46 eV (400K), respectively [42].

As shown in Fig. 6(a)-(c), all the dissolution mechanisms are endothermic at 300K. In Fig. 6(a), ΔG of one $H_2$ molecule incorporation (black line) is the smallest, corresponding to the lowest equilibrium $P_{H2}$ of $1.20 \times 10^{11}$ Pa, which indicates that penetration of $H_2$ is feasible as $P_{H2}$ larger than $1.20 \times 10^{11}$ Pa. Fig. 6(b) and Fig. 6(c) reveal that M2 with 24 $H_2$ molecules (blue line) and M3 with 32 H (red line) have the lowest $\Delta G$ with equilibrium $P_{H2}$ of $1.07 \times 10^{28}$ Pa and $2.86 \times 10^{17}$ Pa, respectively. In Fig. 6(a) and (c), the hydrogen solubility would increase from one $H_2$ to three $H_2$ molecules and 32H to 48H driven by the increasing $P_{H2}$. While in Fig. 6(b), the preferred solubility will not change with $P_{H2}$ increase.

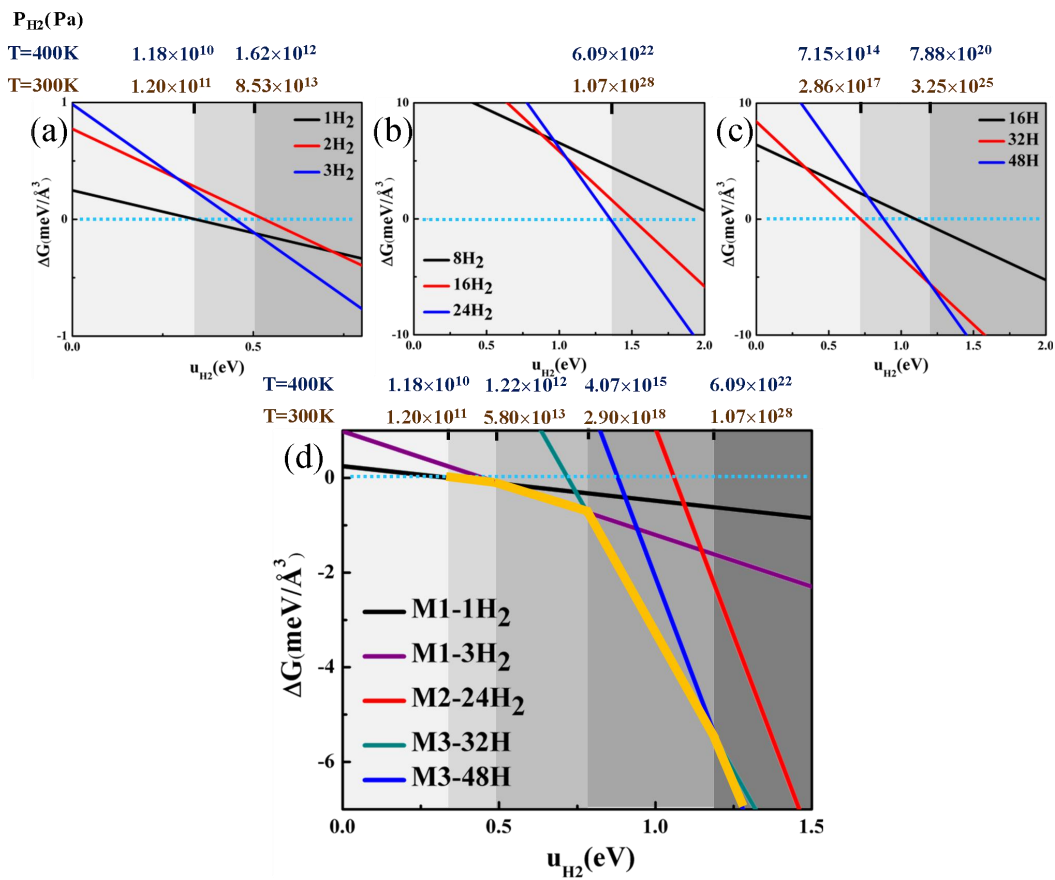

Fig. 6 Generic free energy plot for three kind of H dissolution mechanisms (a) M1, (b) M2 and (c) M3 in equilibrium with $P_{H2}$, all the preferred dissolution cases are shown in (d). $P_{H2}$ at critical point of $\Delta G < 0$ and dissolution mechanism changes are labeled at 300K (in brown) and 400K (in navy).

In order to investigate the transition of dissolution mechanism driven by $P_{H2}$ at 300K, the preferred dissolution states in M1, M2 and M3 are drawn together in Fig. 6(d). We can see that (i) when $P_{H2} < 1.20 \times 10^{11}$ Pa, all the dissolution mechanisms are infeasible; (ii) for $1.20 \times 10^{11}$ Pa $< P_{H2} < 5.80 \times 10^{13}$ Pa, M1 with one $H_2$ appears; (iii) for $5.80 \times 10^{13}$ Pa $< P_{H2} < 2.90 \times 10^{18}$ Pa, M1 with three $H_2$ appears; (iv) if $2.90 \times 10^{18}$

Pa < $P_{H2}$ < $1.07 \times 10^{28}$ Pa, M3 with mixed H and $H_2$ (32H) occurs; (v) as $P_{H2}$ > $1.07 \times 10^{28}$ Pa, mixed H and $H_2$ (48H) appears. The dissolution mechanism transition from M1 to M3 with the increasing $P_{H2}$ is indicated by the yellow line in Fig. 6(d). The influence of temperature on dissolution mechanisms is also discussed supposing that the structure and solubility of dissolved hydrogen will not change distinctly within 400K. At 400K, the equilibrium $P_{H2}$ for M1 is $1.18 \times 10^{10}$ Pa, and $P_{H2}$ for M1 → M3 transition is $4.07 \times 10^{15}$ Pa, which is three orders of magnitude smaller than those at 300K. Therefore, increasing $P_{H2}$ and temperature can promote the subsequent polymolecular and (H + $H_2$) mixed dissolution of hydrogen in α-$Pu_2O_3$.

## 4. Conclusions

Within the DFT+*U*-D3 schemes, the microscopic behaviors of hydrogen in defect-free α-$Pu_2O_3$ are systematically studied by the first-principles calculations and *ab initio* thermodynamic method. Our theoretical study reveals that hydrogen incorporation, diffusion and dissolution behaviors in α-$Pu_2O_3$ are all endothermic. In α-$Pu_2O_3$, H would spontaneously recombine as $H_2$ which is the most stable existence state, owing to the porosity and irreducibility of α-$Pu_2O_3$. And the diffusion of H and $H_2$ in α-$Pu_2O_3$ are both feasible. The high pressure $P_{H2}$ and temperature facilitate hydrogen dissolution from single $H_2$ molecule to (H + $H_2$) mixed dissolution in α-$Pu_2O_3$. Based on a series of theoretical studies, we conclude that the main hydrogen behaviors in defect-free α-$Pu_2O_3$ are not involved with chemical reactions. We also clarify that defect-free α-$Pu_2O_3$ can not catalyze $H_2$ dissociation, and the abundant oxygen vacancies increase the hydrogen solubility. These results provide strong implications to the interpretation of the complex induction period of Pu hydriding.

**Acknowledgments**

This work was supported by National Science Foundation of China, No. U1630250 and No. 11904027. Science Challenge Project, No. TZ2016004 and No. TZ2018002. And partially supported by Foundations for Development of Science and Technology